\documentstyle [11pt]{article}
\begin{document}

 \newcommand \be {\begin{equation}}
\newcommand \ee {\end{equation}}
 \newcommand \ba {\begin{eqnarray}}
\newcommand \ea {\end{eqnarray}}
\def\oppropto{\mathop{\propto}}
\def\operarrow{\mathop{\longrightarrow}}
\def\opsimeq{\mathop{\simeq}}

 \newcommand \s {\sigma}
\newcommand \G {\Gamma}
\newcommand \la {\lambda}
\newcommand \La {\Lambda}
 \newcommand \al {\alpha}
 \newcommand \NE {\not=}
 \newcommand \N {{\cal N}}

\title{\bf RANDOM WALK ON THE BETHE LATTICE \\
AND \\  HYPERBOLIC BROWNIAN MOTION}

\vskip 3 true cm

\author{  C\'ecile Monthus$^1$
 and Christophe Texier$^2$ }

\vskip 2 true cm

\date{ $^1$
 Service de Physique Th\'eorique,
C. E.  Saclay,  
91191 Gif-sur-Yvette C\'edex, FRANCE  \\
$^2$ Division de Physique Th\'eorique\thanks{ 
Unit\'e de Recherche des Universit\'es Paris 6 et Paris 11 associ\'ee au CNRS },  
IPN B\^atiment 100, 
91406 ORSAY C\'edex , FRANCE }

\vskip 3 true cm

\maketitle

\begin{abstract}

{We give the exact solution to the problem of a random walk on the Bethe lattice
through a mapping on an asymmetric random walk on the half-line.
We also study the continuous limit of this model, and
discuss in detail the relation between the random walk on the Bethe lattice 
and Brownian motion on a space of constant negative curvature.  }
\end{abstract}

\vskip 2 true cm

\vfill

\noindent Electronic addresses : \hfill \break
 monthus@amoco.saclay.cea.fr  \hfill \break
texier@ipncls.in2p3.fr

\vskip 2 true cm

\noindent  \hfill Septembre 1995

\newpage

\section {Introduction}

The Bethe lattice, or the infinite Cayley tree, presents a hierarchical
structure that greatlty simplifies some problems of statistical physics.
It has therefore been widely used to obtain analytical results for problems
that are otherwise intractable on Euclidean lattices. 
 The physical relevance of these results 
 is that the Bethe lattice is supposed to represent some mean field limit 
of Euclidean lattices of very large dimensions.
 
The tree structure is also directly relevant to characterize the phase space
hierarchical structure of some disordered systems like spin glasses 
 \cite{Mez86}, where this structure plays an essential role in the understanding
 of the slow
dynamics properties of these systems \cite{Pal85} \cite{Sib87} \cite{Bou95}.  

Random walks on Bethe lattices have already been studied either
for their own interest \cite{Hug82} \cite{Cas89} \cite{Gia95}, or in the
context of polymer physics \cite{Hel83} \cite{Nec87}, where counting
unentangled loops of a polymer in an array of obstacles is equivalent to counting
random walks on Cayley trees.
 In this paper, we give the exact solution to this problem of random walk on the Bethe lattice
and discuss the correspondance with Brownian motion on a space 
of constant negative curvature. 

The paper is organized as follows.
In section \ref{RWBethe}, we consider a mapping between random walk on the Bethe lattice
and a biased random walk on the half-line with reflection at the origin. 
We derive asymptotic expressions at large time through
 a simple argument, and write
the exact explicit solution using old results of Mark Kac \cite{Kac47}.  
Section \ref{BMwithdrift} is devoted to biased Brownian motion on the half-line 
with a reflective barrier at the origin, which represents the continuous limit 
of the previous discrete model. Finally in Section \ref{HyperbolicBM}, we discuss
the relation between the Bethe lattice and hyperbolic geometry. We compare the properties of
Brownian motion on a space of constant negative curvature 
to the results obtained in the previous sections.

\section {Random walk on the Bethe lattice and
biased random walk on the half-chain} 
\label{RWBethe}

Hughes and al. \cite{Hug82} have pointed out that a random walk on the Bethe lattice
can be mapped onto a biased one-dimensional lattice walk with a reflecting barrier 
at the origin, provided one only considers the
distance to the origin on the Bethe lattice as a function of time. 
In this section, we first rederive this correspondence to fix our notations.
We then use the generating function formalism and a Tauberian theorem
to obtain the large time asymptotic behavior of the probability of being at
any given distance from the origin at time t. This result generalizes
a previous one concerning the asymptotic behavior of the probability of
returning to the origin at time t
\cite{Hel83} \cite{Cas89} \cite{Gia95}. We finally use an old result of Kac \cite{Kac47}
to write the full solution of the problem of random walk on the Bethe lattice in
an exact closed form.

\subsection{Master Equation}

We consider a random walk on the Bethe lattice of coordination number $z>2$.
We take the origin of the random walk as the origin of coordinates, and
call generation $n$ the set of all the $z(z-1)^{n-1}$ 
sites distant from the origin by
$n\geq1$ bonds (see Fig.1). 

The random walk is defined as follows.
At each time step,
the particle jumps with probability $\left({1 \over z}\right)$
 to any of its $z$ nearest neighbors.
Therefore, the particle goes further from the origin with probability
 $\left({z-1 \over z}\right)$,
and goes closer to the origin with probability $\left({1 \over z}\right)$.
The corresponding Master Equation for the probability $f_{\tau}(n)$ 
of being on any site of generation n  after $\tau$ time steps reads
\be
f_{\tau+1}(n) = \left(1-{1\over z}\right)f_{\tau}(n-1)
 +{1\over z} f_{\tau}(n+1) \quad \hbox{for} \quad n\geq 2
\label{RWn}
\ee
This has to be supplemented by specific rules
 at the boundary sites $n=1$ and $n=0$
\ba 
&f_{\tau+1}(1) &=f_{\tau}(0) +{1\over z} f_{\tau} (2) \label{RW1}  \\
&f_{\tau+1}(0) &= {1\over z} f_{\tau}(1)  \label{RW0}
\ea
and by the intitial condition at $\tau=0$
\be
f_{\tau=0}(n) = \delta_{n,0} 
\label{RWI}
\ee
The normalization reads $\displaystyle \sum_{n=0}^{\infty}
 f_{\tau}(n) =1$ for any $\tau$. 

The random walk on the Bethe lattice can therefore be described as an 
asymmetric one-dimensional random walk on the half-line ${n\geq 0}$ with 
 a reflecting barrier at the origin $n=0$ \cite{Hug82}.
One may then use standard techniques for one-dimensional random walks.

\subsection{Asymptotic behavior at large time}

It is convenient to introduce the generating function 
\be
F_{n} (\lambda) = \sum_{\tau=0}^{\infty} \lambda^{\tau} f_{\tau} (n)
\label{gene}
\ee
The master equation (\ref{RWn}) for $f_{\tau} (n)$  is then converted into
 the recurrence relation 
\be
F_{n+1} (\lambda) = {z \over\lambda }   F_{n} (\lambda)
 - (z-1)  F_{n-1} (\lambda) \quad \hbox{for} \quad n\geq 2
\label{RecGen}
\ee
and the boundary rules (\ref{RW1}-\ref{RW0}) 
together with the initial condition (\ref{RWI})
give the relations  
\ba 
&F_{2} (\lambda) &={z \over\lambda }   F_{1} (\lambda)
 - z  F_{0} (\lambda) \\
&F_{1} (\lambda) &={z \over\lambda } \big[  F_{0} (\lambda) -1 \big]
\label{Rec012}
\ea
Two linearly independant solutions of (\ref{RecGen}) are $\left[ r(\lambda) \right]^n$
 and $\left[ r'(\lambda) \right]^n$ 
where $r(\lambda)$ and  $r'(\lambda)$ are the solutions of
the characteristic equation
\be
r^2-{z \over\lambda }r +(z-1)=0
\label{carac}
\ee
As we must only keep the regular solution in the limit $\lambda \to 0$ (\ref{gene}),
 we get for $n \geq 2$
\be
F_{n} (\lambda) = \left[ r(\lambda) \right]^{n-1} F_{1} (\lambda) \quad \hbox{with} \quad 
r(\lambda) = {z \over {2\lambda} } \bigg[1- \sqrt{1-{4 \over z^2}(z-1)\lambda^2} \bigg]
\label{Recsol}
\ee
The two first terms $F_{1} (\lambda)$ and $F_{0} (\lambda)$ are then determined
by (\ref{Rec012})
\ba 
&F_{0} (\lambda)  &={ {2 \left({ {z-1} \over z} \right)} \over
{ \left({ {z-2} \over z} \right)+ \sqrt{1-{4 \over z^2}(z-1)\lambda^2} } }  
 \\
&F_{1} (\lambda) &={z \over\lambda } \left({ {1-\sqrt{1-{4 \over z^2}(z-1)\lambda^2} }
 \over {{z-2 \over z}+ \sqrt{1-{4 \over z^2}(z-1)\lambda^2}}} \right)
\label{Recsol01}
\ea
The probabilities $f_{\tau} (n)$ may then in principle be obtained by expanding
the previous expressions (\ref{Recsol}-\ref{Recsol01}) in powers of $\lambda$
 (\ref{gene}). This has been done for the function $F_{0} (\lambda)$ to
obtain $f_{\tau} (0)$ in an explicit form involving hypergeometric functions
\cite{Hel83} \cite{Gia95}. 
This has not been done for the general case $n >0$, where it is preferable
to use some other method to get an exact explicit expression of $f_{\tau} (n)$
 for any n (See next paragraph \ref{Exactsolution}).
 
However, the asymptotic behavior at large time of $f_{\tau} (n)$ 
for any fixed $n$ may be easily found
from the generating function using a Tauberian theorem.

Introducing the notations $\nu=\displaystyle {z-2 \over z}$ and 
$\alpha=\displaystyle{2\over z} \sqrt{z-1}$,
the generating function reads for $n \geq 1$
\be
F_{n} (\lambda) = 2 \left[
{   {z \over{ 2 \lambda}}   \left(
{1- \sqrt{1-(\alpha\lambda)^2}}  \right) } \right]^n
{ 1\over {\nu+\sqrt{1-(\alpha\lambda)^2} } } 
= g_{n}(\alpha \lambda)
\label{Recsolbis}
\ee
with the auxiliary function
\be
g_{n}(x)= 2 
 \left[{ {z \alpha \over{ 2 x}} \left({1- \sqrt{1-x^2}} \right) } \right]^n
{ 1\over {\nu+\sqrt{1-x^2} } }
\label{auxi}
\ee
that admits the power series expansion
\be
g_{n}(x)= \sum_{ \{k \geq 0 ; (k-n) \hbox{even}\}} c_k(n) \ x^k
\label{auxdev}
\ee

Since we have the simple relation $f_{\tau}(n)=c_{\tau}(n) \alpha^{\tau}$, the
asymptotic behavior of the probability $f_{\tau}(n)$ at large time $\tau$
 is directly related to the behavior of the coefficients  
 $ c_k(n)$ for large order $k$. The latter can readily be obtained from the
expansion of $g_n$ (\ref{auxi}) near $x=1$
\be
g_{n}(e^{-E}) \opsimeq_{E \to 0^+} { 2 \over \nu} \left({z \alpha \over 2 }\right)^n
\left[1-\left(n+{ 1 \over \nu}\right) \sqrt{2E}+O(E) \right]
\label{expdir}
\ee
Using the power series expression (\ref{auxdev}), we get
\be
 \sum_{ \{k \geq 0 ; (k-n) \hbox{even}\}} c_k(n) \left( 1-e^{-kE} \right)
 \opsimeq_{E \to 0^+} 
{ 2 \over \nu} \left(n+{ 1 \over \nu}\right)
 \left({z \alpha \over 2 }\right)^n
 \sqrt{2E} +O(E) 
\label{identif}
\ee
The non-analyticity in $\sqrt{E}$ implies that the coefficients $c_k(n)$ 
present the following algebraic decay at large order $k$ for $(k-n)$ even
\ba
 c_k(n) \opsimeq_{k \to \infty} {A(n) \over k^{3 \over 2}}  \\ 
\label{alge}
\ea
The prefactor $A(n)$ can also be obtained from (\ref{identif}) by converting 
the sum into an integral 
\be
A(n) 
= { 2 \over \nu}  \left(n+{ 1 \over \nu}\right) \left({z \alpha \over 2 }\right)^n
 {2^{3\over 2} \over { \displaystyle \int_0^{\infty} du \  
\left( { {1-e^{-u}} \over u^{3\over2}} \right)   }   }
= {2^{3\over 2} \over \sqrt{\pi} }{ (1+n\nu) \over {\nu^2 }} 
 \left({z \alpha \over 2 }\right)^n
\label{pref}
\ee
 We finally get for $(k-n)$ even 
\be
f_{\tau}(n)=c_{\tau}(n) \ \alpha^{\tau} 
 \opsimeq_{\tau \to \infty} A(n) \  {\alpha^{\tau} \over {\tau}^{3 \over 2}} \  
\label{asytaub}
\ee
or more explicitly
\be
f_{\tau}(n) 
 \opsimeq_{\tau \to \infty}
  {2^{3\over 2} \over \sqrt{\pi} } \ \left({z \over {z-2}}\right)^2
 \left(1+n{z-2 \over z}\right)  
 \left( \sqrt{z-1 }\right)^n \  {1 \over {\tau}^{3 \over 2}} \  
e^{\displaystyle -\tau \ln \left({z \over {2\sqrt{z-1}}}
\right)}
\label{asytaubex}
\ee
This formula valid for $n \geq 1$ generalizes the result obtained previously
 for the case $n=0$ \cite{Hel83} \cite{Cas89} \cite{Gia95}
\be
f_{\tau}(0) 
 \opsimeq_{\tau \to \infty}
  {2^{3\over 2} \over \sqrt{\pi} } \ {{z(z-1)} \over {(z-2)^2}}
  \  {1 \over {\tau}^{3 \over 2}} \  
e^{\displaystyle -\tau \ln \left({z \over {2\sqrt{z-1}}}
\right)}
\label{asytaubex0}
\ee

\subsection{Exact expression for $f_{\tau}(n)$}  
\label{Exactsolution}

In \cite{Kac47}, Kac studied the asymmetric random walk (\ref{RWn})
with the reflective boundary condition (\ref{RW1}-\ref{RW0}), but in the domain of
negative drift $1<z < 2$. Using our notations, the explicit expression obtained
 in this paper for the initial condition (\ref{RWI}) reads 
\ba
\displaystyle f_{\tau}^{(0<z<2)} (n) &= \left({2-z}  \right) 
\bigg[ \delta_{n,0}+ z \left(z-1\right)^{n-1} \left(1-\delta_{n,0} \right) \bigg]
 \bigg[{ {1+(-1)^{n+\tau}} \over 2} \bigg]  \label{disc} \\
 &+\displaystyle {2 \over \pi} 
\bigg[ \left({ {z-1} \over z} \right) \delta_{n,0}
+  \left(z-1\right)^{n \over 2} \left(1-\delta_{n,0} \right) \bigg] 
 \left( { {2 \sqrt{z-1}} \over z} \right)^{\tau}  I(n,\tau)   \label{cont}
\ea
where
\be
I(n,\tau) =  \big[1+(-1)^{n+\tau} \big] \int_0^{\pi\over 2} d\theta \left( \cos\theta \right)^{\tau} 
{ {\tan^2\theta } \over { \left({{z-2} \over z} \right)^2+\tan^2\theta} }\ 
\left[ \cos n\theta + \left({{z-2} \over z} \right) { {\sin n \theta } \over {\sin \theta}}
\cos \theta \right] 
\label{Intau}
\ee
The first term (\ref{disc}) can be
easily recovered as the normalized "stationary" (invariant under the time
translation $\tau \to \tau+2$)
solution of (\ref{RWn}) in the case of negative drift $1<z < 2$.

For the case of positive drift $z>2$, which we consider in this paper, there is no normalizable
stationary solution. The exact expression for $f_{\tau}$ describing the random walk on 
the Bethe lattice therefore only contains the second term (\ref{cont}) 
\be
f_{\tau}^{(z>2)} (n) =  {2 \over \pi} 
\bigg[ \left({ {z-1} \over z} \right) \delta_{n,0}
+  \left(z-1\right)^{n \over 2} \left(1-\delta_{n,0} \right) \bigg] 
 \left( { {2 \sqrt{z-1}} \over z} \right)^{\tau}  I(n,\tau)  
\label{ExactBethe}
\ee

For the particular case $n=0$, the integral $I(0,\tau)$ (\ref{Intau}) may be computed
through the change of variable $x=\cos^2\theta$  
\be
I(0,2\tau) = 2\int_0^{\pi \over 2} d\theta \left( \cos\theta \right)^{2\tau} 
{ {\tan^2\theta } \over { \left({{z-2} \over z} \right)^2+\tan^2\theta} }\ 
=\int_0^1 dx \ { {  x^{\tau-1} \  (1-x)^{1\over 2}  } 
\over  { 1- \left(  {{4 (z-1)} \over z^2} \right) x }  }
\label{I0tau}
\ee
which is a standard integral representation of the hypergeometric function
\be
I(0,2\tau) =B\left( \tau+{1 \over 2},{3 \over 2}\right) \ 
F\left (1,\tau+{1 \over 2},\tau+2,{ {4 (z-1)} \over z^2} \right) 
\label{I0hyperg}
\ee
We therefore recover the expression given in \cite{Hel83} \cite{Gia95} 
\be
f_{\tau}^{(z>2)} (0) =  \left({ {z-1} \over z} \right) 
 \left( { { \sqrt{z-1}} \over z} \right)^{2\tau} \ 
{ {\Gamma(2\tau+1)}  \over { \Gamma(\tau+1) \ \Gamma(\tau+2)   } }  \ 
F\left (1,\tau+{1 \over 2},\tau+2,{ {4 (z-1)} \over z^2} \right)
\label{ExactBethe0}
\ee

We may also recover from the expression (\ref{ExactBethe}) the asymptotic expression
at large time given previously in (\ref{asytaubex}). Indeed for large $\tau$, the integral
$I(n,\tau)$ (\ref{Intau}) is dominated by the vicinity of $\theta=0$. The estimation of
the leading order 
\ba
I(n,\tau) &\displaystyle \opsimeq_{\tau \to \infty} \big[1+(-1)^{n+\tau} \big] 
\int_0^{\infty} d\theta \ e^{ -\tau {\theta^2 \over 2}}  \ \theta^2 \ 
\left[ 1 + \left({{z-2} \over z} \right) n \right] \\
&\displaystyle = \left[ { {1+(-1)^{n+\tau} } \over 2} \right] \sqrt{2 \pi} \left({z \over {z-2}} \right)^2 
\left[ 1 + \left({{z-2} \over z} \right) n \right] {1 \over{ \tau^{3 \over 2}}}
\ea
gives back (\ref{asytaubex}).

\section { Biased Brownian motion on the half-line}
\label{BMwithdrift}

We consider the continuum counterpart of the biased random walk described
 in section (\ref{RWn}). 
The probability density $P(x,t)$ for a Brownian particle of diffusion constant $D$
submitted to a constant positive drift $\mu$ satisfies the Fokker-Planck equation
\be
{\partial P \over \partial t} =  {\partial  \over \partial x} \left(
D{\partial P \over \partial x} -\mu P \right)
\label{FP}
\ee
It must be supplemented by the reflection condition at $x=0$ 
\be
 \left(D{\partial P \over \partial x} -\mu P \right) (x=0,t) =0
\label{FPbc}
\ee
to ensure conservation of probability, and by the initial condition
\be
P (x,t \to 0^+) =\delta(x)
\label{FPic}
\ee

\subsection{Expression of the probability density $P(x,t)$}

It is convenient to perform the transformation
\be
P (x,t ) = e^{ {\mu x \over {2D}}} \  \psi(x,t)
\label{FP-SC}
\ee
in order to cast the Fokker-Planck equation (\ref{FP}) into
 the Euclidean Schr\"odinger equation
\be
{\partial \psi \over \partial t} = -  H \psi
\label{SC}
\ee
The corresponding Hamiltonian is a free one up to a constant shift 
\be
H=-D{ d^2 \over dx^2} +{\mu^2 \over {4D}}
\label{H}
\ee
The boundary condition and the initial condition read respectively
\be
 \left({\partial \psi \over \partial x} - {\mu \over {2D}}  \psi \right) (x=0,t) =0
\label{SC-bc}
\ee
and
\be
\psi (x,t \to 0^+) =\delta(x)
\label{SC-ic}
\ee
Using plane waves, it is easy to construct an orthonormal basis 
of eigenvectors $\{\psi_k(x),k\in[0,+\infty[\}$
\be
H \psi_{k}=\left(Dk^2+{\mu^2 \over {4D}}\right) \psi_{k}
\label{evp}
\ee
satisfying the boundary condition (\ref{SC-bc})
\be
\psi_k(x)={1 \over {\sqrt{2 \pi}}} 
\left( e^{-ikx} -{{\mu+2iDk} \over {\mu-2iDk} } \  e^{ikx} \right)
\label{psik}
\ee
The Green function $\psi(x,t)$ solution of (\ref{SC}-\ref{SC-bc}-\ref{SC-ic})
may be expanded onto this orthonormal basis
\be
\psi(x,t) =<x \vert e^{-tH} \vert 0> =
\int_0^{+\infty} dk \ \psi_k(x) \ \psi_k^*(0) \ 
e^{  -t \left( D k^2+ {\mu^2 \over {4D}}  \right) }
\label{Green}
\ee

As a side remark, let us mention that in the case of negative drift $\mu <0$,
in addition to the continuous spectrum (\ref{psik}), 
there is also a bound state solution of zero energy which must be added to 
the expansion (\ref{Green}). This zero-energy state corresponds to the aforementioned 
stationary solution of Kac the discret case (\ref{disc}). 

However, in the case of positive drift $\mu>0$ that we consider here, there is
no stationary normalizable solution, and the spectrum in purely continuous
(\ref{Green}).
The resulting integral
\be
\psi(x,t) ={1 \over \pi}\int_{-\infty}^{+\infty} dk \ \left( { { k^2-ik{\mu \over {2D}}} \over
 { k^2+{\mu^2 \over {4D^2}} } } \right)
\ e^{ikx} \ e^{  -t \left(D k^2+ {\mu^2 \over {4D}}\right)}
\label{ink }
\ee
can be computed through the following trick
\be
\psi(x,t) = e^{\displaystyle -t  {\mu^2 \over {4D}}} 
\left( -{\partial^2 \over \partial x^2}
-{\mu \over {2D}} {\partial \over \partial x} \right) I(x)
\label{deri }
\ee
The integral $I(x)$, being the Fourier transform of the product 
of a Gaussian by a Lorentzian, may be written as a convolution
$$
I(x) = {1 \over \pi} \int_{-\infty}^{+\infty} dk \ \ e^{\displaystyle ikx} 
\left( {1 \over { k^2+{\mu^2 \over {4D^2}}} } \right)
\ e^{-t D k^2} 
$$
\be 
={D \over \mu} \int_{-\infty}^{+\infty} dy \ e^{\displaystyle - {\mu \over {2D}} \vert x-y \vert}
\ {1 \over \sqrt{\pi D t  }} \  e^{\displaystyle - {y^2 \over 4Dt}}
\label{conv }
\ee
Going back to the probability density (\ref{FP-SC}), we finally obtain
\be
P(x,t)={ { e^{\displaystyle -t  {\mu^2 \over {4D}}}   } \over \sqrt{\pi D t  }}
\ e^{ \displaystyle x{\mu \over { 2D} }} \ \left[ e^{ \displaystyle - {x^2 \over 4Dt}} - {\mu \over {2D}} 
\int_x^{\infty} dy \  e^{\displaystyle - {\mu \over { 2D}} (y-x )} \ 
 e^{\displaystyle - {y^2 \over 4Dt}} \right]
\label{P1 }
\ee
It is useful to transform the previous expression into
\be
P(x,t) = {    { e^{ \displaystyle -t  {\mu^2 \over {4D}}}   } 
\over {2 \sqrt{\pi} (D t)^{3 \over 2}  }} \ e^{\displaystyle {\mu \over {2D}} x} \  
\int_x^{\infty} dy \ y \ e^{\displaystyle - {\mu \over {2D}} (y-x) } \ 
  e^{\displaystyle - {y^2 \over 4Dt}}
\label{P2}
\ee
The asymptotic behavior at large time $t$ immediately follows  
\ba
  P(x,t ) & \displaystyle \opsimeq_{t \to \infty} 
 {   {  e^{ \displaystyle -t  {\mu^2 \over {4D}}}      } 
\over {2 \sqrt{\pi} (D t)^{3 \over 2}  }  } \ 
 e^{\displaystyle {\mu \over D}x}  \ 
\int_x^{\infty} dy \ y \ e^{\displaystyle - { {\mu \over {2D}} y} } \\
&=\displaystyle 2    
 \sqrt{D \over \pi} \ {1 \over \mu^2} \    
\left( 1+ x{ \mu  \over {2D} }  \right)
\ e^{ \displaystyle {\mu \over {2D}} x} \  
\  {   1 \over { t^{3 \over 2}  }} \ e^{\displaystyle -t  {\mu^2 \over {4D}}}
\label{tinfty}
\ea
Comparison with the corresponding result for the
discrete case (\ref{asytaubex})
\be
f_{\tau}(n) 
 \opsimeq_{\tau \to \infty}
  {2^{3\over 2} \over \sqrt{\pi} }\left({z \over {z-2}}\right)^2
 \left(1+n{z-2 \over z}\right)  
\  e^{\displaystyle n \ln \sqrt{z-1} } \  {1 \over {\tau}^{3 \over 2}} \  
e^{\displaystyle -\tau \ln \left({z \over {2\sqrt{z-1}}}
\right)}
\label{asytaubexbis}
\ee
shows that even if the two expressions behave qualitatively the
same as $t^{-{3 \over 2}} e^{-At}$ as a function of time, and as
$ \left(1+nC\right) e^{n B } $
as a function of space, there is no direct detailed correspondance at this stage.
If one naively tries to identify $x=n$ and $t=\tau$, then there is no way to find
two functions $\mu(z)$ and $D(z)$ to make equations (\ref{tinfty}) and
(\ref{asytaubexbis}) identical. The relation between the random walk on the Bethe lattice 
and the continuous model (\ref{FP})
studied in this section therefore needs further clarification.

\subsection{Discussion of the continuous limit}

To define properly the continuous limit of the random walk on the Bethe lattice, 
we must introduce some unit length $\Delta x$ and some unit time $\Delta t$ 
into the Master equation (\ref{RWn})
\be
f(x,t+\Delta t) = \left(1-{1\over z}\right) f(x- \Delta x,t)
 +{1\over z} f(x+\Delta x,t) 
\label{RWunits}
\ee
The Taylor expansion
\be
{\partial f \over \partial t} = 
\left( { {(\Delta x)^2} \over {2 \Delta t}} \right) \  {\partial^2 f  \over \partial x^2} 
- \left( { {z-2} \over z} \ { {\Delta x} \over { \Delta t}} \right) \ 
{\partial f \over \partial x}
\label{FPeq}
\ee
gives the Fokker-Planck equation (\ref{FP}) in the limit 
\be
\Delta x \to 0^+   \qquad ; \qquad
\Delta t \to 0^+ \qquad ; \qquad
z \to 2^+ 
\label{limit}
\ee
with 
\be
{ {(\Delta x)^2} \over {2 \Delta t}} = D   \qquad \hbox{and} \qquad
{ {z-2} \over 2} \ { {\Delta x} \over { \Delta t}} = \mu
\label{limitrel}
\ee
where $D$ and $\mu$ are fixed.
This continuous limit therefore requires an analytic 
continuation to non-integer coordination number $z$ 
in order to take the limit $z \to 2^+$. 
With the prescription (\ref{limit}-\ref{limitrel}), the result (\ref{asytaubex}) 
for the discrete case indeed corresponds to the result
 (\ref{tinfty}) of the continuous case.

Let us now write for the discrete case the
analogue of the transformations that we performed for the continuous case
in order to establish some correspondance with the work of Clark et al. \cite{Cla80}.
The analogue of the transformation (\ref{FP-SC}) to make the walk (\ref{RWn})
symmetric 
\be
f_{\tau} (n) = e^{ \displaystyle n \ln \sqrt{z-1} } \ \tilde{f}_{\tau} (n)
\label{discsym}
\ee
and the extraction of the shift factor analogue to the shift present in (\ref{H})
\be
\tilde{f}_{\tau} (n) = e^{\displaystyle - \tau \ln \left({ z \over {2 \sqrt{z-1}}} \right) } 
\  p_{\tau} (n)
\label{shift}
\ee
transform the drifted random walk with reflection at the origin 
(\ref{RWn}-\ref{RW1}-\ref{RW0}) into a symmetric random walk with a partial
absorption at the origin
\ba
&p_{\tau+1}(n) &= {1 \over 2}\  p_{\tau}(n-1)
 +{1\over 2}\  p_{\tau}(n+1) \quad \hbox{for} \quad n\geq 2  \label{pRWn}  \\
&p_{\tau+1}(1) &= \gamma \ p_{\tau}(0) +{1\over 2} p_{\tau} (2) \label{pRW1} \hfill \\
&p_{\tau+1}(0) &= {1\over 2} \ p_{\tau}(1)  \label{pRW0} \hfill
\ea
where $\gamma = {z \over {2 (z-1) }}$ denotes the reflection coefficient at the origin. 
The continuous limit of this random walk with partial absorption at the origin
is indeed a quantum mechanical free problem on the half-line with some
mixed boundary condition (\ref{SC-bc}). The detailed discussion of this point contained
in \cite{Cla80} is equivalent to our approach (see \ref{limit}-\ref{limitrel}).

The meaning of the continuous model (\ref{FP})
to represent the random walk on the Bethe lattice is now clear. 
We will now present another continuous model where the effective drift
 comes from the geometry of the underlying space itself.

\section { Bethe lattices and Hyperbolic geometry}
\label{HyperbolicBM}

A link between Cayley trees and hyperbolic geometry 
has already been emphasized in \cite{Mos82} \cite{Kho85} through the introduction 
of tessalations. Indeed, for any integer $z \geq 3$, one can construct a
 tessalation of the Poincar\'e upper half-plane with polygons with $z$ sides.
Joining the centers of adjacent polygons gives a Bethe lattice 
of coordination number $z$.
In the following, we will not use this approach, but rather
 establish some intrinsic link between Bethe lattices and 
 spaces of constant negative curvature without introducing any tessalation. 

\subsection {Hyperbolic geometry}
  
An N-dimensional Riemannian manifold of constant negative Gaussian
 curvature $K= -{1 \over a^2}$ may be described by the metric 
\be
ds^2_{\em N}=dr^2+a^2 \sinh^2{r \over a}\  d\sigma^2_{\em N-1}
\label{metric}
\ee
where $r \in [0,+\infty]$ measures the distance to the origin, and where
$d\sigma^2_{\em N-1}$ denotes the metric of the unit-sphere $S_{\em N-1}$.
For example, $d\sigma^2_1=d\theta^2$ is the metric of the unit circle in terms
of polar angle $\theta$, and $d\sigma^2_2=d\theta^2+ \sin\theta^2 d\phi^2$
is the metric of the unit sphere $S_2$
in terms of the spherical angles $(\theta,\phi)$.
The volume element $dV$ is covariantly defined as
\be
dV_{\em N} = \left( a \sinh{r \over a} \right)^{N-1} dr \ d\Omega_{\em N-1} 
\label{volume}
\ee
where $d\Omega_{\em N-1}$ is the surface element of the unit-sphere $S_{\em N-1}$; for example
$d\Omega_{1}=d\theta$ and $d\Omega_{2}= \sin\theta d\theta d\phi$.

In particular, the volume of a ball of radius $R$ reads
\be
V_{\em N}(R)= \Omega_{\em N-1} \ a^N \ \int_0^{R \over a} dx \  \left(  \sinh{x} \right)^{N-1}
\label{VolR}
\ee
where $\Omega_{\em N-1}$ is the total surface of the unit-sphere $S_{\em N-1}$; for example
$\Omega_1=2\pi$ and $\Omega_2=4\pi$. This volume has the important property to grow exponentially
with the distance $R$ 
\be
V_{\em N}(R) \oppropto_{R \gg a} e^{\displaystyle R \left(  {N-1 \over a} \right) }
\label{Volexp}
\ee
in contrast with the power dependance $R^N$ in Euclidean space of dimension $N$.

In the same way, the number of sites on the Bethe lattice up to generation $n$ 
\be
{\cal N}(n) = 1+z+z(z-1)+z(z-1)^2+ \cdots+ z(z-1)^{n-1} =
{ {z(z-1)^{n} -2 } \over {z-2}}
\label{Bethen}
\ee
 grows exponentially with the number $n$ of generations
\be
{\cal N}(n) \oppropto_{n \gg 1 } e^{\displaystyle n \ln (z-1) }
\label{Betheexp}
\ee
This is why the Bethe lattice is often considered
to be like a Euclidean lattice of infinite dimension. But it is certainly more interesting to 
relate it to hyperbolic geometry which presents the same property (\ref{Volexp}).
More precisely, we may introduce some lattice spacing $\Delta r$ on the Cayley tree 
so that the sites of generation $n$ are at distance $r=n \Delta r$ from the origin.
Then the exponential dependence of (\ref{Betheexp}) corresponds to (\ref{Volexp})
in the following continuous limit of the Bethe lattice 
\be
\Delta r \to 0^+   \qquad ; \qquad
n \to \infty \qquad ; \qquad
z \to 2^+  
\label{treelimit}
\ee
with 
\be
n \Delta r =r  \qquad \hbox{and} \qquad
{ {z-2} \over {\Delta r}}  = {N-1 \over a}
\label{Bethelimitrel}
\ee
where $N$ and $a$ are fixed. 

\subsection {Hyperbolic Brownian motion}

The radial part $\Delta_r$ of the Laplace operator on the N-dimensional Riemannian 
manifold of constant negative Gaussian
 curvature defined by the metric (\ref{metric}) reads  
\be
\Delta_r= {1 \over \left(\sinh {r \over a} \right)^{N-1}} \ 
 {{\partial} \over {\partial r}}
\left[ \left(\sinh {r \over a} \right)^{N-1}  {{\partial} \over {\partial r}} 
 \right] 
\label{Laplacian}
\ee
On this manifold, free Brownian motion starting from the origin is defined by the
 diffusion equation for the Green's function $G_{\em N} (r,t)$
\be
{ {\partial G_{\em N}} \over {\partial t}} = D \Delta_r  G_{\em N}  
\label{diffusion}
\ee 
and the initial condition 
\be
G_{\em N} (r,t) \operarrow_{t \to 0^+} \delta(r) {1 \over { r \ \Omega_{\em N-1} }}
\label{inco}
\ee
The normalization of the Green function $G_{\em N} (r,t)$
 then reads for any time t
\be
1=\int dS  \ G_{\em N} (r,t) = \Omega_{\em N-1}  \int_{0}^{+\infty} dr \  
\left(a \sinh{r \over a}\right)^{N-1}  G_{\em N} (r,t) 
\label{norma}
\ee
The solution reads respectively in two and three dimensions 
\be
G_2 (r,t) = { 
 {e^{\displaystyle -{Dt \over {4a^2}} }} \over {4 \sqrt{2} a (\pi D t)^{3\over 2} }}
 \ \int_r^{\infty} dy \ { {y \ e^{\displaystyle -{y^2 \over {4Dt}}} } \over 
{\sqrt{\cosh\left({y \over a}\right) -\cosh\left({r \over a}\right) } }}
\label{GreenHyp2}
\ee
and
\be
G_3 (r,t) = 
{  {  e^{\displaystyle -{Dt \over {a^2}} }} \over {8  a (\pi D t)^{3\over 2} }}
 \ { {r \ e^{\displaystyle -{r^2 \over {4Dt}}} } \over 
{\sinh \displaystyle {r \over a}}  }
\label{GreenHyp3}
\ee

 Consider now the probability density $P_{\em N}(r,t)$ to be at time t at a distance r 
from the origin
\be
P_t(r) = S_{\em N-1}  \left(a \sinh{r \over a}\right)^{N-1}  G_{\em N} (r,t) 
\label{probdens}
\ee
normalized with respect to the flat measure $dr $ (\ref{norma})
\be
1 = \int_{0}^{+\infty} dr \   P_{\em N} (r,t)
\label{normprob}
\ee
This probability density satisfies the Fokker-Planck equation (\ref{diffusion})
\be
{ {\partial P} \over {\partial t}}  =
   {{\partial} \over {\partial r}}
\left[ 
 D{{\partial P} \over {\partial r}} - D\left({N-1 \over a}\right) 
\coth\left({r \over a}\right) P \right] 
\label{FPhyp}
\ee
and the initial condition at $t=0$ (\ref{inco})
\be
P_{\em N} (r,t) \operarrow_{t \to 0^+} \delta(r) 
\label{FPhypinco}
\ee
The solution reads respectively in two and three dimensions (\ref{GreenHyp2}-\ref{GreenHyp3})
\be
P_2 (r,t) = {  {e^{\displaystyle -{Dt \over {4a^2}} }  \over
 {2 \sqrt{2\pi}  ( D t)^{3\over 2} }} \ \sinh\left({r \over a}\right) } \ 
 \int_r^{\infty} dy \ { {y \ e^{\displaystyle -{y^2 \over {4Dt}}} } \over 
{\sqrt{\cosh\left({y \over a}\right) -\cosh\left({r \over a}\right) } }}
\label{probsol2}
\ee
\be
P_3 (r,t) =a {  {  e^{\displaystyle -{Dt \over {a^2}} }} \over {2  \sqrt{\pi} ( D t)^{3\over 2} }}
 \ r \ \sinh{r \over a} \ e^{\displaystyle -{r^2 \over {4Dt}}}   
\label{probsol3}
\ee

At large distance from the origin $r \gg a$, the Fokker-Planck equation 
(\ref{FPhyp}) corresponds
to a one-dimensional diffusion with a constant drift $\mu= D{ {N-1} \over a}$ 
(\ref{FP})
\be
{ {\partial P} \over {\partial t}}  \sim    {{\partial} \over {\partial r}}
\left[ 
D {{\partial P} \over {\partial r}} - D\left({ {N-1} \over a} \right) P \right] 
\label{FPhypasy}
\ee
One may check that this identification of the effective radial constant drift 
$\mu= D{ {N-1} \over a}$
at large distance on the hyperbolic space is entirely consistent with the
two continuous limits of the Bethe lattice previously defined in
 (\ref{limit}-\ref{limitrel}) and (\ref{treelimit}-\ref{Bethelimitrel}).

The solutions (\ref{probsol2}-\ref{probsol3}) in two and three dimensions
read approximatively at large distance
\be
P_{2} (r,t) \opsimeq_{r \gg a} {  {e^{\displaystyle-{Dt \over {4a^2}} }  } \over
 {4 \sqrt{\pi}  ( D t)^{3\over 2} }} \  e^{\displaystyle {r \over {2a}} } \ 
 \int_r^{\infty} dy \ y \ e^{-\displaystyle {{y-r} \over {2a} }}
\    e^{\displaystyle -{y^2 \over {4Dt} }}  
\label{FPsolasy2}
\ee
\be
P_{3} (r,t) \opsimeq_{r \gg a}  a \ 
{  {  e^{\displaystyle -{Dt \over {a^2}} }} \over {4  \sqrt{\pi} ( D t)^{3\over 2} }}
 \ r \ e^{\displaystyle{ r \over a}}\  e^{\displaystyle -{r^2 \over {4Dt}}} 
\label{FPsolasy3}
\ee
to be compared with (\ref{P2}).
These expressions at large time
\be
P_{2} (r\gg a,t \to \infty) \simeq { a \over {2 \sqrt{\pi}   D^{3\over 2} }} 
\ r \  e^{\displaystyle {r \over {2a}} } 
\  { 1 \over   t^{3\over 2} } \    e^{\displaystyle-{Dt \over {4a^2}} } 
\label{FPsolasy2t}
\ee
\be
P_{3} (r,t) \opsimeq_{r \gg a}  { a \over {4 \sqrt{\pi}   D^{3\over 2} }} 
 \ r \ e^{\displaystyle{ r \over a}}\  
 { 1 \over   t^{3\over 2} } \    e^{\displaystyle-{Dt \over {a^2}} } 
\label{FPsolasy3t}
\ee
are the exact analog of expression (\ref{tinfty}) at large distance
\be
  P(x \gg {2D \over \mu} ,t \to \infty ) =  
 \ {1 \over {2 \mu  \sqrt{ \pi D}}  } \    
x \ e^{ \displaystyle {\mu \over {2D}} x} \  
\  {   1 \over { t^{3 \over 2} }} \ e^{\displaystyle -t  {\mu^2 \over {4D}}}
\label{tinftyx}
\ee
for $\mu= D{ {N-1} \over a}$. This confirms the equivalence with biased 
Brownian motion on the half-line.

\section{Conclusion}

The exact solution (\ref{ExactBethe}) to the homogenous random walk on the Bethe lattice
was obtained through a mapping onto a biased one-dimensional random walk on the half-line
that describes the "radial" dynamics on the tree. This mapping to a one-dimensional
lattice can of course be extended to inhomogenous random walks on the Bethe lattice
that still preserve the radial invariance on the tree 
\cite{Hug82} \cite{Cas90} \cite{Asl91}, 
but does not hold any longer as soon as two sites belonging to the same generation 
are no more equivalent. This is in particular the case when one considers
 disordered problems on the Bethe lattice.   
This is why the relation with hyperbolic geometry that we presented
in section \ref{HyperbolicBM} is much more profound. On a 
Riemanian manifold of constant negative curvature, we have at our disposal
not only the radial coordinate that corresponds to the generation number
on the Bethe lattice, but also angular coordinates, that correspond
to the "degree of freedom" inside a given generation of the tree.
In this paper, we have discussed in detail the radial correspondance, but it would
certainly be interesting to consider also more precisely the angular part, 
and to study in particular
the continuous limit of some disordered systems on the Bethe lattice.

\vskip 1.5 true cm

\leftline{\bf Acknowledgements}

\vskip 0.5 true cm

We wish to thank Alain Comtet for many helpful discussions and for his remarks on the manuscript.

\vskip 2 true cm

\end{document}